\documentstyle[11pt,epsf]{article}

\newcommand{\f}{\varphi}
\newcommand{\C}{V(\f)}
 
\title{
\vspace{0cm}\large\bf
On the Equation of State for Scalar Field
}
\author{
Alexander S. Silbergleit\\
Gravity Probe B, W.W.Hansen Experimental Physics Laboratory,\\
Stanford University, Stanford, CA 940305-4085, USA.\\
 e-mail: {\it gleit@relgyro.stanford.edu}
}

\date{~}

\begin{document}

\maketitle

\begin{abstract}
\noindent We consider Friedmann cosmologies with minimally coupled scalar field. Exact solutions are found, many of them elementary, for which the scalar field energy density, $\rho_f$, and pressure, $p_f$, obey the equation of state (EOS) $p_f=w_f\rho_f$. For any constant $|w_f|<1$ there exists a two-parameter family of potentials allowing for such solutions; the range includes, in particular, the quintessence ($-1<w_f<0$) and `dust' ($w_f=0$). The potentials are monotonic and behave either as a power or as an exponent for large values of the field. For a class of potentials satisfying certain inequalities involving their first and second logarithmic derivatives, the EOS holds in which $w_f=w_f(\f)$ varies with the field slowly, as compared to the potential.


\end{abstract}

\noindent{\bf 1. Introduction} 
\vskip1.5mm
\noindent Quintessence is a dynamic energy with negative pressure-to-density ratio \cite{cald}. The idea of quintessence provides new `degrees of freedom' in cosmology \cite{cald,2}, extends the variety of modern field models to include extreme forms of energy \cite{cald,3}, and may stimulate a better understanding of the fundamental problem of interplay between gravity and field theories \cite{4,w}. Quintessence is also shown to give rise to spacetimes with new interesting properties \cite{gon0, css}.

However, one needs to find out what kind of physical substance obeys the quintessence EOS, that is, to model quintessence, in some way. It was common understanding from the very beginning that a scalar field could act as quintessence; nevertheless, it should be clearly demosntrated.

The EOS for the scalar field has been intensively studied in \cite{zl1} for the so called tracking cosmological solutions introduced in \cite{zl2}, and some classes of potentials allowing for the field EOS were described. In this letter we examine this question for any cosmological solutions. We first give an explicit description of all potentials allowing for solutions with scalar field satifying the linear EOS with constant parameter $w_f$. Then, by way of generalization, we derive the conditions on the potential which provide the EOS where the parameter varies slowly with the field, $w_f=w_f(\f)$.
\vskip1.5mm
\noindent{\bf 2. Exact solutions with scalar field satisfying linear EOS}
\vskip1.5mm

\noindent We consider Friedmann universe described by the Friedmann-Robertson-Walker metric,
\begin{equation}\label{metr}
ds^2= - dt^2 + a^2(t)\left( \frac{dr^2}{1-kr^2} + r^2d\Omega^2\right) \; ,
\end{equation}
where $k= -1, 0$, or $1$, according to whether the universe is open, flat, or closed ($G=c=1$). In the presence of matter of the density, $\rho$, and  pressure, $p$, and a minimally coupled scalar field, $\f$, with the potential $\C\geq0$, the evolution equations can be written in the form:
\begin{eqnarray}
3\,\frac{\dot{a}^2}{a^2}= 8 \pi  \rho + 8 \pi  \rho_f - 3\, \frac{k}{a^2}\; ;
\label{h} \\
\dot{\rho} = -\,3\,\frac{ \dot{a}}{a}\, (\rho + p)\; ;
\label{e} \\
\dot{\rho_f} = -\,3\,\frac{ \dot{a}}{a}\, (\rho_f + p_f)\; . 
\label{f}
\end{eqnarray}
The proper definitions of the scalar field energy density, $\rho_f$, and pressure, $p_f$,  are: 
\begin{equation}\label{pf}
  8\pi\rho_f\equiv\dot{\f}^2 + V ,\qquad
8\pi p_f\equiv\dot{\f}^2 - V\; .
\end{equation}
The first of them is implied by equation (\ref{h}), the second one is validated by the very equation (\ref{f}), written as the field energy conservation equation, identical in its form with the conservation law (\ref{e}) for matter. Also, the expression for the acceleration, following from eqs. (\ref{h})--(\ref{e}) and (\ref{pf}), reads:
\begin{eqnarray}
3\,\ddot{a}/{a}=-4\pi\left(\rho+3p\right)+V-2\dot{\f}^2\equiv
-4\pi\left[(\rho+3p)+(\rho_{f}+3p_{f})\right)]
\;; \label{ac}
\end{eqnarray}
the density and pressure of both matter and field are involved here uniformly.

We assume the linear equation of state for the matter, $p=w\rho,\quad w={\rm const}$. Our main interest is concerned with the `normal' matter with $w\geq0$; however, all the following results hold for any $w\geq-1$. Using the EOS, equation (\ref{e}) immediately integrates to express the density via the scale factor ($C>0$ is an arbitrary constant):
\begin{equation}\label{dens}
 8\pi\rho=C/a^{3(1+w)}\; .
\end{equation}

Global dynamics of cosmological expansion governed by the system (\ref{h})--(\ref{f}) was  described in \cite{ss}. In this paper we are looking for special solutions subject to an additional constraint, namely, a linear EOS for the scalar field, $p_f=w_f\rho_f$.
First, we assume $w_f$ constant,  and look for potentials which would allow solutions satisfying such EOS. Introducing expressions (\ref{pf}) converts the EOS into an equivalent relation,
\begin{equation}
\dot{\f}^{2}=\frac{1+w_f}{1-w_f}\,\C\; , 
\label{fdot}
\end{equation}
showing that the acceptable range of the EOS parameter is $|w_f|<1$. Evidently, the EOS of the Zel'dovich superstiff fluid, $w_f=1$, holds only for the free field, $\C\equiv0$; the vacuum EOS ($w_f=-1$) is only valid for the constant field $\f_c$ such that, by (\ref{f}), $V^{'}(\f_c)=0$. This is the classical case of a universe with the cosmological constant, studied recently in detail in \cite{s}. 

For any given $\C$ and $w_f$, equation (\ref{fdot}) completely specifies the time evolution of the scalar field. On the other hand, the EOS of the scalar field allows for an immediate integration of the field energy conservation equation (\ref{f}), in exactly the same fashion as with the equation (\ref{e}) for the matter, so that
\begin{equation}\label{densf}
 8\pi\rho_f=C_f/a^{3(1+w_f)}\; ;
\end{equation}
here  $C_f>0 $ is another arbitrary constant of motion. Combining this and (\ref{fdot}), one relates the potential (as a function of time for the solution under investigation) to the scale factor,
\begin{equation}\label{pot}
 V=(1-w_f)\,C_f/2a^{3(1+w_f)}\; .
\end{equation}

At this point, only the equation (\ref{h}), of the whole system (\ref{h})--(\ref{f}), remains to be satisfied. By (\ref{dens}) and (\ref{densf}), it transforms into the first order equation for the scale factor (the chosen positive sign corresponds to expansion):
\begin{equation}
{\dot{a}}= \left(C/3a^{1+3w} + C_f/3a^{1+3w_f} - {k}\right)^{1/2}\; .
\label{sfact}
\end{equation} 
 
This, in fact, completes the solution: integration of equation (\ref{sfact}) gives the scale factor as a function of time, next, relation (\ref{pot}) provides the potential, also as a function of time. Scalar field $\f(t)$ is then determined by integrating a known function of time:
\begin{equation}
\dot{\f}(t)=\pm\sqrt{\frac{1+w_f}{1-w_f}\,V}=
\pm\sqrt{\frac{(1+w_f)\,C_f}{\,\,2\,[a(t)]^{3(1+w_f)}}}\; . 
\label{scfi}
\end{equation} 
Formula (\ref{scfi}) shows that $\f=\f(t)$ is a monotonic function of time, so that the inverse to  function $t=t(\f)$ is well defined. Hence, the potential can be finally determined as the function of the field, 
$\C=(1-w_f)\,C_f/2\left[a(t(\f))\right]^{3(1+w_f)}$, 
according to the expression (\ref{pot}).

Using the EOS  for both matter and scalar field, and then (\ref{dens}) and (\ref{densf}), the acceleration equation (\ref{ac}) for the found solution becomes
\begin{equation}
3\,\ddot{a}/{a}= 
-(1/2)\left[(1+3w)C /a^{3(1+w)}+(1+3w_f)C_f/a^{3(1+w_f)}\right]\; .
 \label{ac2}
\end{equation}
Even if the matter is `normal', $w\geq0$, but the field acts as quintessence with $-1<w_f<-1/3$, the expansion clearly accelerates, at least at large enoguh values of $a$. Also, if $w_f\approx0$, then the scalar field can play the role of dark matter in the universe (whether the scalar field can clamp to `normal' matter, as the dark matter seemingly does, requires separate investigation).

The found solution contains, along with $C,\,C_f$, two more arbitrary constants coming from integration of  (\ref{sfact}) and (\ref{scfi}). Both of them are not significant, since they just shift the values of time and scalar field. Thus we have demonstrated that, given any $|w_f|<1$ (and $w>-1$), a two--parameter family of potentials $V=V(\varphi, C, C_f)$ and the corresponding cosmological solution exists, for which the EOS of the scalar field holds throughout the whole evolution. If $-1<w_f<0$, then the scalar field acts exactly as quintessence, and, consequently, for $-1<w_f<-1/3$  the expansion of the universe accelerates, at least at its late stages.
\vfill\eject

\noindent{\bf 3. Explicit description of the potential}
\vskip1.5mm

\noindent The above derivation lacks a direct formula relating the potential with the scalar field, $V=\C$. The formula is not difficult to establish: one just writes the evolution equation (\ref{h}) not in terms of the scale factor, as in (\ref{sfact}), but rather in terms of the potential. Indeed, from (\ref{pot}), one has $a=[(1-w_f)\,C_f/2V]^{1/3(1+w_f)}$; differentiating this in time using (\ref{fdot}), one obtains also a formula for $\dot{a}/{a}$ through $V$ and $dV/d\f$. 
Replacing all the terms in the evolution equation (\ref{sfact}) with their ready expressions via $V$ results in the first order differential equation for $\C$:
\begin{equation}
Q(V,w_f)\,{dV}/\,V = 
\mp\sqrt{6\,(1+w_f)}\, d\f \; ,
\label{V(fi)}
\end{equation}
where
\begin{eqnarray}\label{Q}
Q(V,w_f)=\left[1+A\,V^{\frac{w-w_f}{1+w_f}}-k\,B\,V^{-\frac{1+3w_f}{3(1+w_f)}}\right]^{-1/2}\\
\label{AB}
A=\left(C/C_f\right)\,\left[(1-w_f)C_f/2\right]^{\frac{w_f-w}{1+w_f}},
\quad B=\left(3/C_f\right)\,\left[(1-w_f)C_f/2\right]^{\frac{1+3w_f}{3\,(1+w_f)}}\;,
\end{eqnarray}
and the signs correspond to the signs in (\ref{scfi}). 

After a straightforward integration of (\ref{V(fi)}) (that leads, in many cases, to an elementary result, see the next section), we arrive at the desired expression for the potential as function of the scalar field, $V=\C$. The dynamics of the field, i. e., the function $\f=\f(t)$, is then determined by integrating the first expression in (\ref{scfi}), and the scale factor evolution is found by its expression through $V(\f(t))$ above. This expression shows that always $V=\infty$ at the Big Bang ($a=0$), and $V\to 0$ when $a\to\infty$; it also demonstrates that the potential is a monotonic function of scalar field. Moreover, equation (\ref{V(fi)}) allows us to specify the behavior of the potential in both limits; the most interesting are the late stages of the cosmological expansion. 

If the EOS parameter satisfies
\begin{equation}
w_f\leq\min\{w,\,-1/3\}\; ,
\label{ex}
\end{equation}
then the potential drops exponentially, $\C\sim \exp(\mp\alpha\f)$. The constant $\alpha>0$ equals to $\sqrt{6(1+w_f)}$ when no equality holds in (\ref{ex}), and is expressed through all the parameters otherwise. So, under the condition (\ref{ex}) the found solutions implement the runaway scenario \cite{w}, with $\f\to\pm\infty,\quad \C\to0$ at large times. Remarkably, all this happens exactly in the case when the scalar field behaves as a `strong' quintessence with the pressuere equally or more negative than the one of the Einstein quitessence, for which $w_f=-1/3$. When condition (\ref{ex}) is invalid, the potential behaves as a power of the scalar field when it is small, $\C\sim \f^{\nu}$; here $\nu>0$ depends only on  $w$ and $w_f$. 

Note, however, that the large--time behavior of the scale factor $a(t)$ and the potential $V(\f(t))$ - as fucntions of time -- is qualitatively the same for all possible parameter values: these functions are, respectively, a positive and negative power of time related by (\ref{pot}). It is only the scalar field, $\f(t)$, that evolves differently at large times, depending on whether condition (\ref{ex}) is true or not: in the first case, the field is logarithmic, in the second case it has a power asymptotics, same as the other two dynamic variables.
\vskip1.5mm
 
\noindent{\bf 4. Examples of closed-form solutions }
\vskip1.5mm

\noindent Several classes of physically meaningful closed--form solutions are described by the same formula for the potential, differing just by the values of the parameters involved. The reason for this is the structure of the key equation (\ref{V(fi)}): there are two terms added to unity under the square root in $Q(V,w_f)$ [see (\ref{Q})]. In all the cases when either of the terms is a constant, the integral of (\ref{V(fi)}) is elementary and, of course, of the same functional form. We give the expression only for the potential [if $\C$ is known, $\f(t)$ and $a(t)$ are found as described in the previous section]. This potential, allowing for solutions with scalar field satisfying a linear EOS, is
\begin{equation}\label{pot2}
\C=V_0\left[\sinh^2m(\f-\f_*)\right]^{\mu},\qquad \f_*={\rm const}\; .
\end{equation}
The expression is valid in the following five cases : 1) flat universe, $k=0$; 2) universe  with scalar field only, $C=0$; 3) same EOS for both matter and field, $w=w_f$; 4) scalar field as Einstein quintessence, $w_f=-1/3$; 5) Einstein quintessence as matter, $w=-1/3$. The cases differ only by the values of $\mu$, $V_0$, and $m$, whose expressions are cumbersome. Thus we show just one example, for the first case of flat universe:
\[
\mu=\frac{1+w_f}{w_f-w},\qquad 
V_0=\frac{(1-w_f)C_f}{2}\,\left(\frac{C}{C_f}\right)^\mu,
\qquad  m=|w_f-w|\,\left[\frac{3}{2\,(1+w_f)}\right]^{1/2}\;.
\]
In each of the above cases it is taken that only one of the five assumptions designating the cases is fulfilled (for instance, in the first case $C\not=0,\,w_f\not=w,\,w_f\not=-1/3$, $w\not=-1/3$, etc.). On the other hand, many combinations of two of those assumptions (such as $k=0$ and $C=0$, $k=0$ and $w=w_f$, etc.) imply  that $Q(V,w_f)$ in the equation (\ref{V(fi)}) becomes a constant. Therefore, the potential proves to be exactly an exponent in every such case ,
\begin{equation}\label{pot3}
\C=V_0\, \exp(\mp\alpha\f)\; ,
\end{equation}
with arbitrary $V_0$ and $\alpha>0$ depending on the parameters. The scalar field then is a logarithmic function of time, which results in the potential proportional to the inverse square of time, $ V(\f(t))\,\sim\,t^{-2}$; the scale factor is also a power, $a(t)\,\sim\,t^{2/3(1+w_f)}$.

Many other closed--form solutions are not related to the degeneracy of $Q(V,w_f)$. To give one example, for pressureless matter ($w=0$) and scalar field obeying the EOS of quintessence with $w_f=-2/3$ in the open universe ($k=-1$), the potential is:
\begin{equation}\label{pot3}
\C=4\,\bar{e}(\f)\,\left\{\left[\bar{e}(\f)-B\right]^2-4A\right\}^{-1},\qquad
\bar{e}(\f)\equiv\exp\left[\pm\sqrt{6(1+w_f)}\,(\f-\f_*)\right] \; ,
\end{equation}
where $A=6C/5C_f^{3},\,\, B=2/5$, according to (\ref{AB}) for this particular case. The stability of cosmological solutions driven by quitenssence with the pressure to density ratio $(-2/3)$ has been recently studied in~\cite{gon}. 
\vskip1.5mm

\noindent{\bf 5. Generalizations. Linear EOS with slowly varying parameter, $w_f(\f)$}
\vskip1.5mm

\noindent Generalization of the above results to the case of arbitrary number, $N$, of non-interacting matter species and arbitrary number, $N_f$, of non-interacting scalar fields $\f_n$ satisfying (different) linear EOS, is straightforward. Such solutions can, in particular, fit the concordant cosmological data, if $N=N_f=2$ and $w^{(1)}=0$ for baryonic matter, $w^{(2)}=1/3$ for radiation, scalar field $\f_1$ plays the role of the dark matter with $w^{(1)}_f=0$, and scalar field $\f_2$ acts as quintessence with $w^{(2)}_f=-1\,\div\,-0.6 $ \cite{conq}.

Also, the EOS of the field holds approximately for long periods of time for the potentials which are small perturbations of those found in the previous sections. However, the scalar field can satisfy the EOS approximately in a much broader range of situations. Suppose the parameter in the EOS for the field is not a true constant, but a function of the field, $w_f=w_f(\f)$, so that the EOS, or the equivalent equation (\ref{fdot}), is no longer a constraint, rather, a definition of a new dynamic variable. The only constraint we now impose is that this variable, $w_f(\f)$, varies slowly with the field as compared to its density $\rho_f$, so that the integration of the field energy conservation equatuion (\ref{f}) resulting in (\ref{densf}) is still approximately valid. By (\ref{fdot}), the condition of slow variation proves to be
\begin{equation}\label{slow1}
\frac{2}{1-w_f}\,\left|\frac{dw_f}{d\f}\right|\ll
\frac{1}{V}\,\left|\frac{dV}{d\f}\right|,\quad
{\rm or\,\, just}\quad
\left|\frac{dw_f}{d\f}\right|\ll\left|\frac{d\ln V}{d\f}\right|\; 
\end{equation}
provided that $w_f(\f)$ is not too close to unity; particularly, the second form of the condition is true for all the negative values of $w_f(\f)$.

Under the condition (\ref{slow1}), all calculations of sections 2 nad 3 hold to lowest order in the slow variation. The key equality (\ref{V(fi)}), which we now write as 
\begin{equation}
w_f=\frac{Q^2(V,w_f)}{6}\,\left(\frac{d\ln V}{d\f}\right)^2\,-\,1\; ,
\label{wf}
\end{equation}
is no longer a differential equation for $\C$, but, for a given potential, a transcendetal equation for $w_f(\f)$, instead. If its proper solution exists, then condition (\ref{slow1}) becomes an inequality which limits the class of potentials in question.

Since this is an important general result, we formulate it accurately. 
Suppose that for some values of $w$, $k=0,\pm 1$ and constants $C,C_f$ involved in $Q$, equation (\ref{wf}) has a solution $w_f=W(V,d\ln V/d\f)$. Then the inequalities 
\begin{equation}
W\left(V,\frac{d\ln V}{d\f}\right)<1,\qquad
\left|\frac{d}{d\f}\,W\left(V,\frac{d\ln V}{d\f}\right)\right|\ll\left|\frac{d\ln V}{d\f}\right|\; 
\label{slow2}
\end{equation}
describe all the potentials for which cosmological scalar field satisfies a linear EOS with the parameter varying slowly with the field. To lowest order in this slow variation, the dynamics of cosmological expansion, i. e., the functions $\f(t)$ and $a(t)$, are found successively from (\ref{scfi}) and (\ref{pot}). If the solution $w_f=W(V,d\ln V/d\f)$ and conditions (\ref{slow2}) are valid not for all the values of the field, but only for some range of it, then, accordingly, the EOS holds only for the corresponding period of cosmological evolution, with the values of $\f$ within this range.

Equation (\ref{wf}) is generally very complicated [recall that $w_f(\f)$ is involved in the powers of $V$, as well as in the `constants' $A$ and $B$, see (\ref{Q}), (\ref{AB})] and should be studied in detail separately. However, it dramatically simplifies for later stages of cosmological expansion under the condition (\ref{ex}), which includes the most interesting case $w\geq0,\,\,w_f<-1/3$ (`normal' matter and `strong' quintessence). Indeed, at the later stages the scale factor is large, $a\gg1$, and, by (\ref{pot}), the potential is small, $V/C_f\ll 1$. Due to this and condition (\ref{ex}), the factor $Q$ gets close to unity, so that (\ref{wf}) turns into an approximate explicit expression for $w_f(\f)$ through the potential,
\begin{equation}
w_f=\frac{1}{6}\,\left(\frac{d\ln V}{d\f}\right)^2\,-\,1\; ,
\label{wf1}
\end{equation}
and conditions (\ref{slow2}) also become explicit and rather simple:
\begin{equation}
\left|\frac{d\ln V}{d\f}\right|<2,\qquad
\left|\frac{d^2\ln V}{d\f^2}\right|\ll 3\qquad\; . 
\label{slow3}
\end{equation}

The set of potentials satisfying (\ref{slow3}) is rather wide. For instance, $\C=V_0(\f^2+b^2)\exp(\alpha\f)$ with $b^2\gg 1$, $|\alpha|<2$,and $\alpha\not=0$ (to make the potential small at large $|\f|$ of the proper sign), satisfies (\ref{slow3}) for all values of $\f$, with the varying part of $w_f$ of the order of $1/b^2$,
\[
w_f=-1+\alpha^2/6 + O(1/b^2)\; .
\]
The factor $(\f^2+b^2)$ in this example can evidently be replaced with any polynomial of an even degree and with no real roots, under a single additional constraint on its coefficients guaranteeing the smallness of its second logarithmic derivative. Next, a ratio of two such polynomials can be taken, in which case the parameter $\alpha$ in the exponent could even be zero, if the degree of the polynomial in the denominator is larger than in the numerator, etc. 

All the above examples correspond to the runaway scenario $\f(t)\to\infty,\,\,V(\f(t))\to0$ at $t\to+\infty$. This is true in general for the solution (\ref{wf}), (\ref{slow3}) at $V/C_f\ll 1$. Indeed, if the scalar field goes to a finite value $\f_0$ at large times, then, by (\ref{pot}), $V(\f_0)=0$. Since the potential is nonnegative, $\f_0$ is the point of its minimum, and $V^{'}(\f_0)=0$.
But then the logarithmic derivative $(\ln V)^{'}$ tends to infinity when $\f(t)\to\f_0$, and equation (\ref{slow2}) cannot be satisfied in the vicinity of $\f_0$. In fact, the EOS of the scalar field proves to be nonlinear in this vicinity.

Thus we have shown that if matter satifies the EOS with $w>-1/3$, and the potential drops at large enough values of the scalar field in such way that inequalities (\ref{slow3}) hold, then the runaway scenario is possible with the field obeying the EOS of a `strong' quintessence, $-1<w_f(\f)<-1/3$, at  later stages of expansion. This regime takes on the earlier in the expansion, the larger the abundance of the field (the ratio $C_f/C$), and the closer its EOS to the EOS of vacuum [$w_f(\f)$ to $-1$] is. So, under conditions (\ref{slow3}) on the potential going to zero at large values of $\f$, scalar field quintessence dominates the expansion of the universe at its later stages in all the runaway solutions, and the expansion is then accelerating.
\vskip1.5mm

\noindent{\bf Acknowledgments}
\vskip1.5mm
\noindent This work was supported by NASA grant NAS 8-39225 to Gravity Probe B. The author is grateful to A.D.Chernin, D.I.Santiago, and R.V.Wagoner for their valuable remarks.
\vfill\eject

\end{document}